\documentclass[twocolumn,showpacs,pre, floatfix]{revtex4}

\ifx\pdfoutput\undefined
\usepackage{graphicx}
\else
\usepackage[pdftex]{graphicx}
\usepackage{epstopdf}
\fi

\usepackage[center]{subfigure}

\begin{document}

\title{Plasticity and Dislocation Dynamics in a Phase Field Crystal Model}

\author{Pak Yuen Chan, Georgios Tsekenis, Jonathan Dantzig, Karin A. Dahmen,
and Nigel Goldenfeld}
\affiliation{Department of Physics, University of Illinois at
Urbana-Champaign, Loomis Laboratory of Physics, 1110 West Green
Street, Urbana, Illinois, 61801-3080.}

\begin{abstract}
The critical dynamics of dislocation avalanches in plastic flow is
examined using a phase field crystal (PFC) model.  In the model,
dislocations are naturally created, without any \textit{ad hoc}
creation rules, by applying a shearing force to the perfectly periodic
ground state.  These dislocations diffuse, interact and annihilate with
one another, forming avalanche events.  By data collapsing the event
energy probability density function for different shearing rates, a
connection to interface depinning dynamics is confirmed. The relevant
critical exponents agree with mean field theory predictions.
\end{abstract}


\pacs{05.10.Cc, 61.72.Cc, 62.20.Dc, 81.16.Rf, 81.40.Jj, 05.70.Ln} \maketitle

Materials yield and deform plastically under large external stress.
While the yield surface and the plastic flow are well described by
various continuum theories\cite{cottrell, mendelson1968pta,
hosford2005mbm}, what happens microscopically during a plastic
deformation is still not fully understood.  On atomic scales, external
stress is carried by localized crystal defects, such as dislocations
and disclinations. They are created under stress and interact with each
other.  Although the properties of individual defects and their
interactions are well known\cite{hirth1982td, nabarro1987tcd}, their collective behavior under
external stress is complicated. It gives rise to scale invariant,
power-law distributed phenomena \cite{zaiser2006sip, weiss2003sfa,
weiss1997aes, miguel2001idf, richeton2005bac, bharathi2001mbs,
bharathi2002cap}, in strong resemblance to the scaling behavior near a
critical point. These phenomena include dislocation slip avalanches of
a broad range of sizes, and acoustic emission with a power law power
spectrum.

Recently, evidence has accumulated that these scaling phenomena reflect
an underlying non-equilibrium critical point\cite{zaiser2006sip,
weiss2003sfa}, i.e. a point where the non-equilibrium steady state
fluctuations are governed by a diverging correlation length
\cite{ODOR04}, as appears to be the case in magnetic
materials\cite{Sethna01} and even turbulence\cite{GOLD06}.  Such a
critical point would exist at the boundary between two distinct
regimes, one of which would be a glassy, activated regime, the other
would be a genuine plastic flow regime.  Up to now, experimental and
computational data have focused on the glassy regime: for example,
Weiss \textit{et al.} have measured the acoustic emission signal from
creep deformation experiments on single crystal ice and found that the
event size distribution follows a power law over $4$
decades\cite{weiss1997aes, weiss2003sfa}.  Miguel \textit{et al.}'s
dislocation dynamics simulations in two dimensions showed event size
distributions following a power law, with a rate-dependent cutoff, over
approximately $2$ decades \cite{miguel2001idf}.  Zaiser \textit{et al.}
have reported a data collapse of the event size distribution with
different external stresses \cite{zaiser2006sip}.

The purpose of this Letter is to approach plastic deformation from the
flow side of the non-equilibrium critical point, manifested in the
strain-rate dependence of the acoustic emission.  Importantly, we are
able to systematically vary the strain-rate in simulations, and
moreover we relate the critical point underlying plastic flow {\it at
finite strain rates} to the scaling  of magnetic domain wall depinning
\cite{Narayan1996sfa,zaiser2006sip,dahmen09}.  We find remarkable
agreement between simulations and analytical mean field theory
predictions of exponents, and in addition are able to show that the
strain-rate data exhibit collapse.  Our results strongly support the
critical point picture of plasticity, and suggest new experiments.
We study dislocation avalanches during plastic flows using the phase
field crystal (PFC) model\cite{Elder02,Elder04}.  This approach is
well-suited to this problem, because it can be performed at finite
temperature, for large systems, and over long time periods. The PFC
model describes the dynamics of the local crystalline density field,
and has been shown to give an excellent account of numerous materials
properties including polycrystalline solidification, vacancy diffusion,
grain growth, grain boundary energetics, epitaxial growth,
fracture\cite{Elder04}, grain coarsening\cite{Singe06},
elasticity\cite{Stefa06}, dislocation annihilation, glides and
climb\cite{Berry06}, as well as vacancy dynamics\cite{Chan09}.  The
model can be derived from density functional
theory and extended to the case of binary alloys\cite{Elder07}. In this
paper, we augment the model to treat external shearing forces by adding
an advective term to the dynamics near the boundary. By adjusting the
shearing force and measuring the resulting avalanche statistics, a data
collapse is obtained, that is consistent with proximity to a domain
wall depinning point at finite temperature
\cite{zaiser2006sip,dahmen09}.

{\it The Model:-\/}
The phase field crystal (PFC) model is given by the free energy density\cite{Elder02,Elder04}
\begin{equation}
f = \frac{\rho}{2}(\nabla ^2 + 1)^2\rho + \frac{r}{2}\rho^2 + \frac{\rho ^4}{4},
\end{equation}
where $r$ is the undercooling and $\rho(\vec{x},t)$ is the local
density.  The dynamics associated with this free energy is
conservative, relaxational and diffusive, and systematically derivable
from density-functional theory\cite{MAJA2007}.  In order to study the
plastic response of the PFC model under shear, we add a shearing term
to the dynamical equation:
\begin{equation}
\frac{\partial^2 \rho}{\partial t^2} + (\beta)
\frac{\partial \rho}{\partial t}
 = (\alpha)^2\nabla ^ 2\frac{\delta F}{\delta \rho} + v(y)\frac{\partial \rho}{\partial x} + \eta,
 \label{eqn_shear_PFC}
\end{equation}
where
\begin{equation}
v(y) = \left\{
\begin{array}{ll}
v_0 e^{-y/\lambda} & \quad\text{for}\quad  0<y<L_y/2 \\
-v_0 e^{-(L_y-y)/\lambda} & \quad\text{for}\quad  L_y/2 < y < L_y
\end{array}
\right.
\end{equation}
is the shearing profile and $v_0$ is the magnitude of the shearing,
$\lambda$ is the penetration depth, $\alpha$ and $\beta$ control the
range and time scale respectively of elastic interactions (phonon excitations)
propagating through the medium\cite{Stefa06}, $F\equiv \int f(\vec{x})
d^dx$ is the total free energy and $\eta$ is the thermal noise
satisfying the fluctuation-dissipation theorem
$\langle \eta(\vec{x},t) \eta(\vec{x}',t') \rangle = -\epsilon \nabla ^2\delta(\vec{x}-\vec{x}')\delta(t-t').$
Here $\epsilon$ is the noise amplitude. It is directly proportional to
the temperature, $k_BT$.  The value of $v_0$ controls the magnitude of
the shearing force; the penetration depth, $\lambda$, controls how deep
the shearing force extends into the material.  In all simulations we
set $\lambda \ll L_y$, so the actual value of $\lambda$ does not affect
our simulation results. The boundary conditions are periodic in $x$ and
fixed at $y=0,L$.  The PFC model we used allows propagating sound
modes\cite{Stefa06} when $\beta=0.9$; as long as $\beta$ is non-zero
and O(1), we do not expect that our scaling results will be sensitive
to its precise value.  Values of $\beta$ that are O(10) would
correspond to very over-damped dynamics, and could model viscoelastic
behavior\cite{Stefa06} that is outside the scope of our work.

One of the advantages of using the PFC model is that we do not have to
impose any ad hoc assumptions about the creation and annihilation of
dislocations.  Recall that in dislocation dynamics simulations,
dislocations are treated as elementary particles and usually only the
far field interaction between dislocations is captured.  When
dislocations get too close to each other (a few atomic spacings), the
highly nonlinear interaction between them is not captured and more
importantly, the annihilation of dislocations is not accounted for.
The standard practice is then to impose some annihilation
rules---dislocations of opposite topological charges annihilate when
they get too close to each other\cite{miguel2001idf}. Similarly,
dislocations have to be created by hand when the local strain is high.
Although these rules are consistent with our physical intuition,
particular ways of implementing them are sometimes difficult to
justify.  However, because the PFC model captures the nonlinear elastic
behavior of a crystal, the interaction between dislocations is
completely captured. In addition, because the PFC model simulates the
atoms in the lattice (the PFC density is periodic in its ground state
with peaks representing atoms and troughs inter-atomic space),
but not the dislocations themselves, creation and
annihilation of them are also naturally captured as collective
excitations of the lattice.  No \textit{ad hoc} rules or assumptions
have to be imposed.

We solved Eq. (\ref{eqn_shear_PFC}) in a $2$D rectangular domain.  The
crystal under shear is initially perfectly triangular.  As the crystal
is sheared, dislocations are created near the fixed boundaries $y=0,L$
where the stress is higher. They then propagate into the bulk.
They interact with each other and form avalanches.  To
quantify the avalanche activity, we calculate the total speed of all
dislocations in the domain,
$\tilde{V}(t)=\sum_{i=1}^{N_{dis}(t)} |\vec{u}_i|$,
where $N_{dis}(t)$ is the number of dislocations in the system at time
$t$ and $\vec{u}_i$ is the velocity of the $i$-th dislocation.  This
measure is similar to the acoustic emission signal in Weiss \textit{et
al.}'s single crystal ice experiments.  As dislocations are generated
and interact with each other in the domain, in addition to the fast
avalanching dynamics, quasi-static structures, such as grain
boundaries, can form.  These slow dynamics should not be measured
because they are really not part of the avalanches.  This leads to the
distinction between fast-moving and slowly-moving dislocations
introduced by Miguel \textit{et al.}\cite{miguel2001idf}.  In essence,
they introduced a cutoff in dislocation speed and measure only
dislocations with speed higher than the cutoff.  In that way, they
tried to retain only the avalanche activities in the acoustic emission
signals.

We employed a different method to tackle this problem.  Instead of
simulating a very large system, with all sorts of dislocation
activities, we simulated a moderate size of system with approximately
$10000$ atoms. For this system size, dislocation avalanches come and
go, \textit{i.e.,} not many dislocations are left in the system after
every avalanche.  As a result, no grain boundaries, or slow dynamics,
is present and we obtain clean avalanche data.  It is fair to mention
that this method severely limits the system size, and thus the
resulting avalanche sizes.  The system size we chose contains
approximately $100$ dislocations in the largest avalanche events.  The
tradeoff, which we exploit, is the cleanness of the avalanche signal
and the speed of the resulting simulations.  Different methods, such as
those we mentioned above, would have to be employed if larger avalanche
sizes are desired.

\begin{figure}
\begin{center}
\includegraphics[width=0.80\columnwidth]{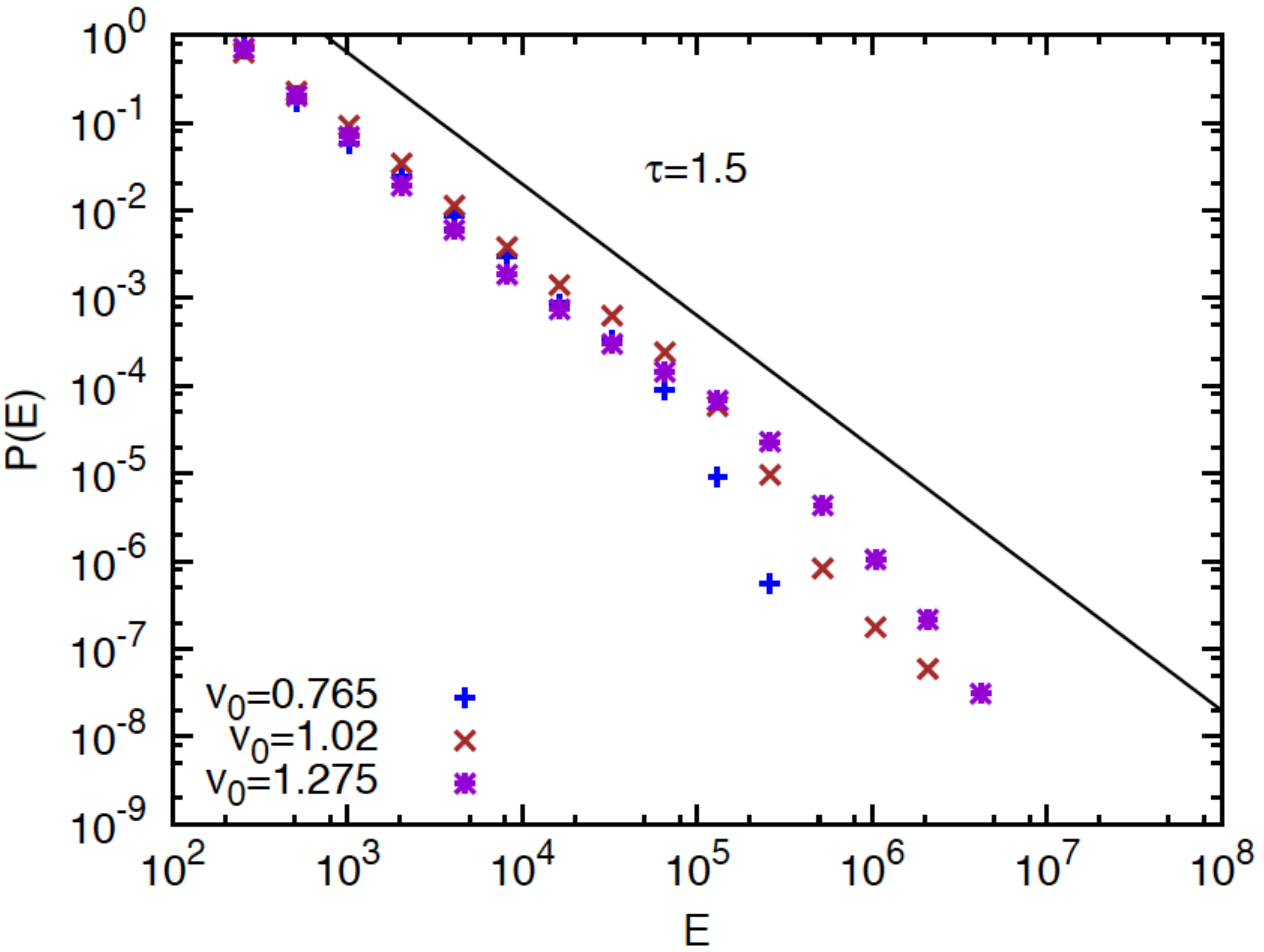}
\caption[The probability distribution of the event energy during
dislocation avalanches.] { (Color online) The probability distribution
of the event energy during dislocation avalanches, for different values
of the shearing rates. } \label{fig_event_dist}
\end{center}
\end{figure}

We count the number of nearest neighbors of each atom, $n_i$, using the
Delaunay triangulation method in computational
geometry\cite{sack2000hcg,orourke1998cgc}.  Because we have $n_i=6$ for
every atom in a perfectly triangular crystal, and because there are no
vacancies introduced in the version of the PFC model simulated here
(vacancies can be introduced into the PFC model by breaking the up-down
symmetry of the PFC free energy, as detailed in \cite{Chan09}), any
atom having $n_i\neq 6$ is sitting next to a dislocation.  Because the
PFC exhibits emergent rigidity in the region of the phase diagram
studied here, these \lq defect atoms' essentially track the locations of
dislocations.  Instead of measuring the total sum of dislocation
speeds, $\tilde{V}(t)$, we then measure the total sum of these defect
atoms' speeds,
\begin{equation}
V(t) = \sum_{i=1}^{N(t)} |\vec{v}_i|,
\end{equation}
where $N(t)$ is the number of defect atoms and $\vec{v}_i$ is the
velocity of defect atom $i$.  Note that the velocity of a defect atom
is not the velocity of any atom in the system, but the velocity of the
dislocation it is tracking.  Because the two measures, $\tilde{V}(t)$
and $V(t)$ are proportional to each other with the proportionality
constant being the mean number of defect atoms sitting next to a
dislocation, we can use the latter for convenience.
\begin{figure}
\begin{center}
\begin{tabular}{c}
\includegraphics[width=0.80\columnwidth]{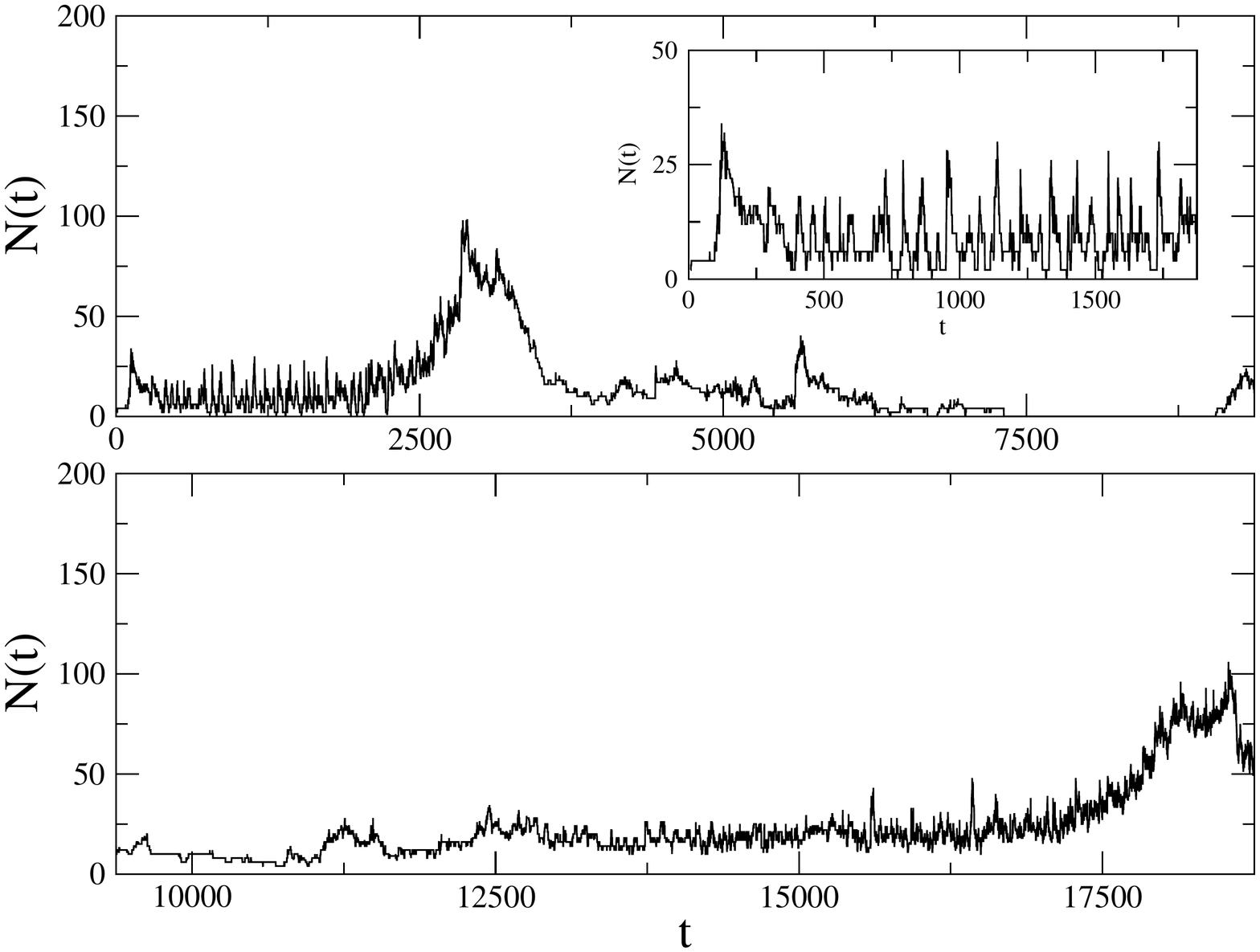}\\
\end{tabular}
\end{center}
\caption[Time dependence of the number of dislocations in a sheared PFC
crystal.] { The number of dislocations in a sheared PFC crystal.
Intermittent events with sizes differing in orders of magnitude is
observed.  Parameters are $(\alpha)^2=255$, $\beta=0.9$, $v_0=1.581$,
$\rho_0 = 0.3$, $\epsilon=1.5$, $\lambda=40.0$ and $r=-0.5$. } \label{fig_num_dis}
\end{figure}
\begin{figure}
\begin{center}
\begin{tabular}{c}
\includegraphics[width=0.80\columnwidth]{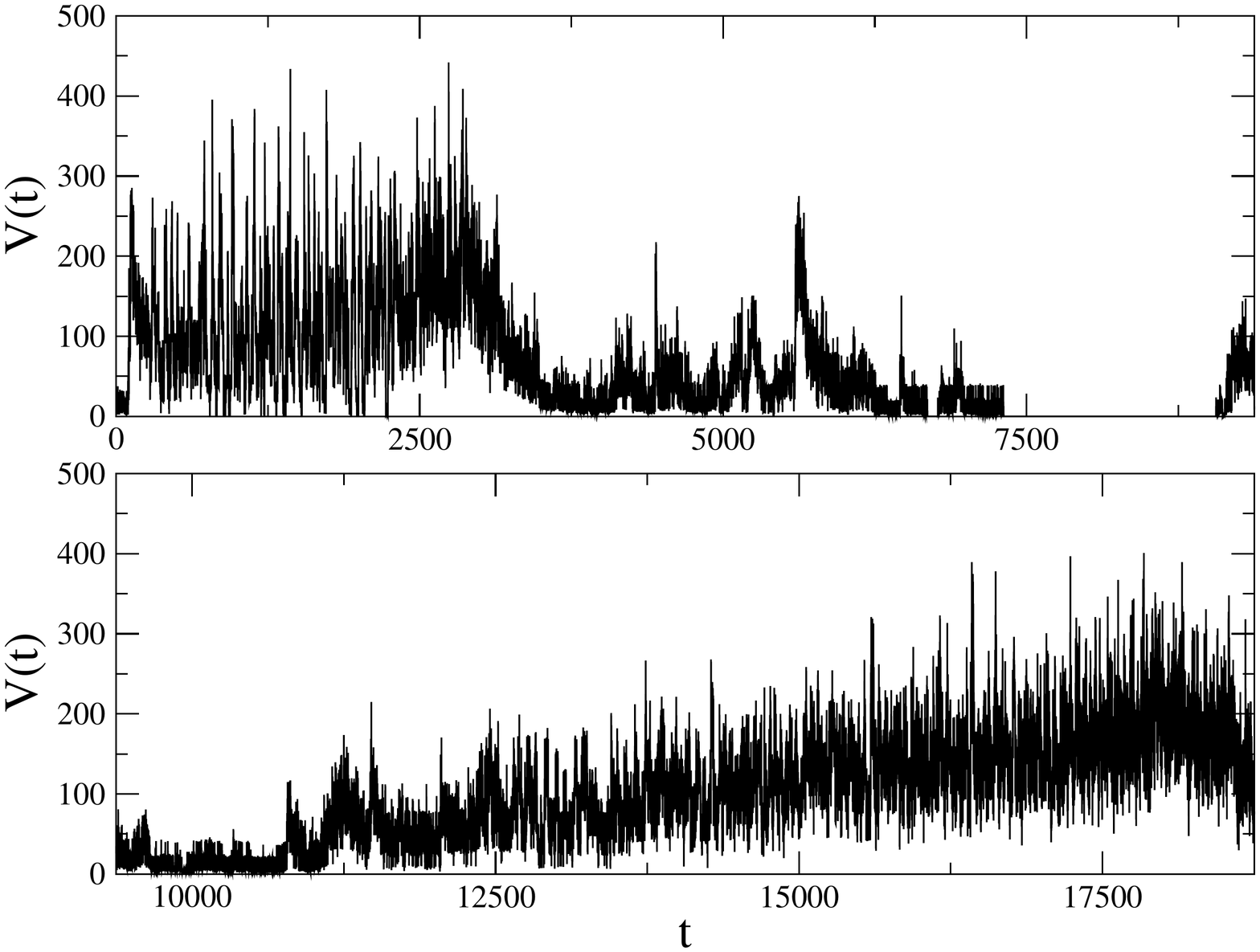}\\
\end{tabular}
\end{center}
\caption[Time dependence of the total speed of defect atoms in a
sheared PFC crystal.] { The total speed of defect atoms in a sheared
PFC crystal.  Intermittent events with sizes differing in orders of
magnitude is observed.  Parameters are $(\alpha)^2=255$, $\beta=0.9$,
$v_0=1.581$, $\rho_0 = 0.3$, $\epsilon=1.5$, $\lambda=40.0$ and $r=-0.5$. } \label{fig_V_dis}
\end{figure}
Fig. \ref{fig_num_dis} shows the typical time dependence of $N(t)$ from
a simulation with parameters $dx=3\pi/8$, $dt=0.025$, $L_x=L_y=512$,
$(\alpha)^2=255$, $\beta=0.9$, $v_0=1.581$, $\rho_0 = 0.3$,
$\epsilon=1.5$, $\lambda=40.0$ and $r=-0.5$.  $N(t)$ changes as
dislocations are being created and annihilated.  There are intermittent
events of creation of dislocations, with number of dislocations
involved ranging from a few to $80$. Fig \ref{fig_V_dis} shows the
acoustic emission signal, $V(t)$, in the same simulation.  Similar to
$N(t)$, the signal ranges from $0$ to $400$, with intermittent pulses
of various sizes.  In order to measure the avalanche event size, we
introduce a cutoff to the signal $V_{cut}=15$ for shear velocity
$v_0=0.765$. The cutoff is increased proportionally to the shear
velocity, and is primarily used to define the avalanche size, rather
than to remove slowly moving dislocations from the analysis. The signal
is then partitioned into individual avalanche events.  The probability
distribution of the event energy,
\begin{equation}
E = \int_{t_{begin}}^{t_{end}} V^2(t) dt,
\label{signal-integrated}
\end{equation}
where $t_{begin}$ and $t_{end}$ are the starting and ending time of the
event respectively, can be measured.  For cutoff values small compared
to the signal (i.e. $V_{cut}=15$, corresponding to the activity of 3-4
dislocations due to thermal creep),
the result is insensitive to the cutoff. For each
shearing rate, at least $10$ different realizations are run to obtain a
statistically meaningful result.  This results in about $8000$
avalanche events for each shearing rate. Fig. \ref{fig_event_dist}
shows the event size distribution for different shearing rates. We find
that the distribution follows a power law for small event sizes and
cuts off at larger sizes, with the cutoff size depending on the
shearing rate.  The data is somewhat noisier towards the end of large
event size because large events are rare.

\begin{figure}
\begin{center}
\includegraphics[width=0.80\columnwidth]{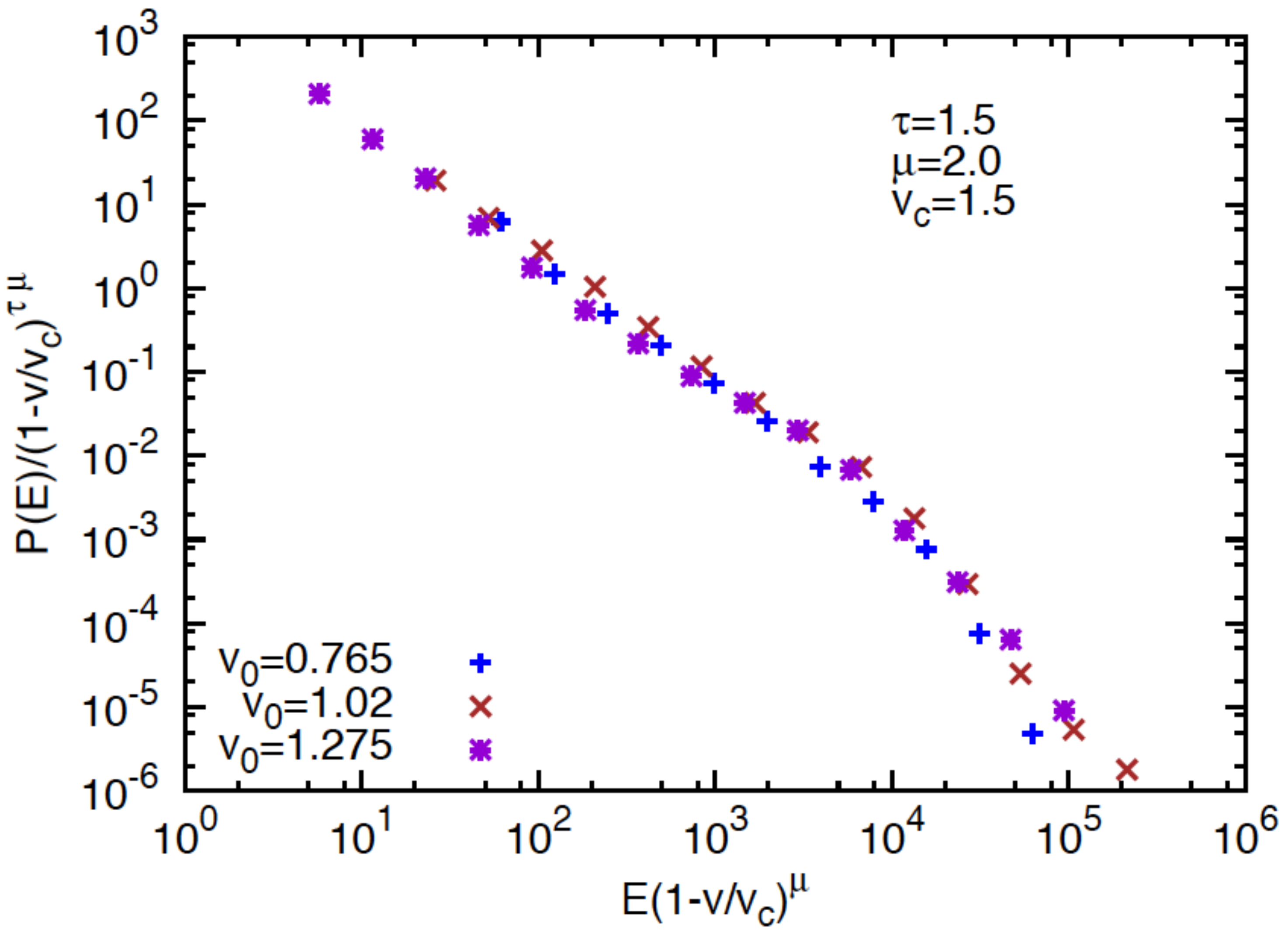}
\caption[Data collapse of the probability distribution of the event
energy during dislocation avalanches.] { (Color online) Data collapse
of the probability distribution of the event energy during dislocation
avalanches, with $\tau=1.5$, $\mu=2$ and $v_c=1.5$.}
\label{fig_event_collapse}
\end{center}
\end{figure}

{\it Scaling Behavior of the avalanches:-\/}
Analogous to the scaling behavior in models of crackling noise
\cite{Sethna01}, we propose that there is a non-equilibrium critical point,
$v_0=v_c$, in the system and we expect, as $E\rightarrow\infty$, the data
around the critical point to collapse in the form
\begin{equation}
P(E,\bar{v}) \sim E^{-\tau} f(E {\bar{v}}^{\mu}),
\label{eqn_prob_asym}
\end{equation}
where $P(E,\bar{v})$ is the probability distribution of event energy,
$E$, $\bar{v}\equiv 1 - v_0/v_c$ is the reduced shearing rate with $v_0$
being the shearing rate and $v_c$ being the critical shearing rate.
$\tau$ and $\mu$ are two critical exponents.  As $\bar{v}\rightarrow
0$, $P(E,\bar{v})$ tends to a power law
$P(E,\bar{v}) \sim E^{-\tau}$.
Fig. \ref{fig_event_collapse} shows an attempt to collapse the data,
using the equivalent scaling form $P(E,\bar{v}) \sim
{\bar{v}}^{\tau \mu} g(E {\bar{v}}^{\mu})$,
with $\tau=1.5$, $\mu=2$ and $v_c=1.5$ and universal
scaling function $g(x)=x^{-\tau}f(x)$ shown in the collapse.
Logarithmic binning is performed and singletons
are ignored to obtain $P(E,\bar{v})$. We obtain a satisfactory data
collapse over $4$ decades, with $\bar{v}$ ranging from $0.15$ to
$0.49$.  As $E\rightarrow 0$, the collapse function $g(x)$
approaches the power law $g(x)\sim x^{-3/2}$, which
agrees   with
$f(x) \rightarrow const$ in Eq. (\ref{eqn_prob_asym}) for
$x \rightarrow 0 $ and $\tau=3/2$.
The collapse constrains the numerical exponent to the range
$\tau=1.5 \pm 0.2$. This agrees with the experimental result
of $\tau=1.5$ \cite{Richeton2006}. Similarly,
for adiabatic stress increase in the pinned
regime, Zaiser finds $\tau=1.4$ \cite{zaiser2006sip}.
Dimiduk reports $\tau=1.5-1.6$ in experiments at fixed compression
stress that leads to shearing \cite{Dimiduk}.
In contrast to the universal distribution of time integrated avalanche
signals discussed here (see eq.(\ref{signal-integrated})),
the signal {\it amplitude} statistics likely depend
on details of the system:
The distribution of energy {\it amplitudes} decays with the exponent $\tau=1.8$
\cite{miguel2001idf} in simulations, and with $\tau=1.6$ in experiments \cite{miguel2001idf}.
The accoustic emission intensity exponent is $\tau=1.8$ in simulations
at fixed stress \cite{Koslowski}.

Studies at adiabatically slow shear rate have suggested analogies to
the depinning transition of magnetic domain walls
\cite{zaiser2006sip,Zapperi98,dahmen09}, although one exponent that may deviate
in simulations is discussed in \cite{miguelbeta}. The exponents found
in our collapse are consistent with the domain wall depinning picture.
As argued previously, mean field theory (MFT) is expected to give exact
scaling results in this case. The MFT values for the exponent $\tau$ is
$\tau=1.5$ \cite{zaiser2006sip,dahmen09}.
The MFT value for the exponent $\mu$ can be calculated from the
MFT of \cite{Narayan1996sfa,Zapperi98,dahmen09}
to be $\mu=2$. These MFT values for the exponents $\tau$ and $\mu$ lead
to a satisfactory collapse of the numerical avalanche size distributions
at different shear rates. At zero temperature the
critical shear rate
is $v_c =0$. Our simulations, however, are performed at finite
temperature $T$. Temperature-induced dislocation creep causes the
critical shear rate to appear to be nonzero, with the apparent $v_c(T)
\rightarrow 0$ as $T \rightarrow 0$. It also causes the scaling
collapse in Fig. \ref{fig_event_collapse} to be less precise than
in zero temperature studies.
We hope to report on a theoretical investigation of the temperature dependence and a
comparison with experiments at finite shear rate in a future
publication.

{\it Acknowledgements:} We thank C. Miguel and J.T.
Uhl for helpful conversations. We acknowledge support from grants
NSF DMR 03-25939 ITR (MCC) (GT and KD) and DOE subcontract 4000076535 (JD).

\bibliographystyle{apsrev}

\bibliography{pfc_plasticity_bib}

\end{document}